\documentclass[a4paper]{jpconf}
\usepackage{graphicx}
\usepackage{amsmath}

\newcommand{\ket}[1]{\big| \,{#1}\, \big> }
\newcommand{\braket}[2]{\big\langle {#1} \big| {#2} \big\rangle}
\newcommand{\matrixe}[3]{\big< \,{#1}\, \big| \,{#2}\, 
  \big| \,{#3}\, \big> }

\renewcommand{\vec}[1]{\mathbf{#1}}
\newcommand{\op}[1]{\hat{#1}}

\begin{document}

\title{Clusters and halos in light nuclei}

\author{Thomas Neff}

\address{GSI Helmholtzzentrum f{\"u}r Schwerionenforschung GmbH,
  Planckstra{\ss}e 1, 64291 Darmstadt, Germany}

\ead{t.neff@gsi.de}

\begin{abstract}
The fermionic molecular dynamics approach uses Gaussian wave packets as single-particle basis states. Many-body basis states are Slater determinants projected on parity, angular momentum and total linear momentum. The wave-packet basis is very flexible -- FMD contains harmonic oscillator shell model and Brink-type cluster states as special cases. The parameters of the wave packets are obtained by variation. A realistic effective interaction derived from the Argonne~V18 interaction by means of the unitary correlation operator method is employed. We discuss the fully microscopic calculation of the $^3$He($\alpha$,$\gamma$)$^7$Be capture reaction within the FMD approach. The model space contains frozen cluster configurations at large distances and polarized configurations in the interaction region. The polarized configurations are essential for a successful description of the $^7$Be bound state properties and for the $S$- and $D$-wave scattering states. The calculated cross section agrees well with recent 
measurements regarding both the absolute normalization and the energy dependence. We also discuss the structure of the cluster states, including the famous Hoyle state, in $^{12}$C. From the two-body densities we conclude that the Hoyle state has a spatially extended triangular $\alpha$-cluster structure, whereas the third $0^+$ state features a chain-like obtuse triangle structure. We also calculate the $N\hbar\Omega$ decomposition of our wave functions to illuminate the challenges of no-core shell model calculations for these cluster states.
\end{abstract}

\section{Introduction}

Cluster and halo states present a challenge for nuclear structure calculations. They are typically found close to thresholds and feature extended tails in the wave functions. In many-body approaches like the no-core shell model where the wave functions are expanded in a harmonic oscillator basis the description of such spatially extended wave functions require huge model spaces. Cluster models on the other hand are by construction well suited to describe the asymptotics of such states. However cluster models are an oversimplification. With realistic interactions it is not possible to obtain reasonable results as polarization effects of the clusters are missing. With the fermionic molecular dynamics approach we attempt to combine the advantages of a microscopic model with those of the cluster model. 

\section{Fermionic Molecular Dynamics}

In the fermionic molecular dynamics (FMD) approach we employ Gaussian wave packets
\begin{equation}
  \label{eq:wavepacket}
  \braket{\vec{x}}{q} = \frac{(\vec{x}-\vec{b})^2}{2 a} \otimes \ket{\chi^\uparrow, \chi^\downarrow} \otimes \ket{\xi}
\end{equation}
as single-particle basis states. The complex parameters $\vec{b}$ encode the mean positions and
momenta of the wave packets and $a$ the widths of the wave packets. The spins can
assume any direction, isospin is $\pm 1$ denoting a proton or a
neutron. Intrinsic many-body basis states are Slater determinants 
\begin{equation}
  \label{eq:sldet}
  \ket{Q} = \mathcal{A} \left\{ \ket{q_1} \otimes \ldots \otimes \ket{q_A} \right\} \: .
\end{equation}
The intrinsic states $\ket{Q}$ reflect deformation or clustering and break the symmetries of the Hamiltonian with respect to parity, rotation and translation. To restore the symmetries the intrinsic basis states are projected on parity, angular momentum and total linear momentum
\begin{equation}
 \ket{Q; J^\pi MK; \vec{P}=0} = \op{P}^\pi \op{P}^J_{MK} \op{P}^{\vec{P}=0} \; \ket{Q} .
\end{equation}
In a full FMD calculation the many-body Hilbert space is spanned by a set of $N$ 
intrinsic basis states $\left\{ \ket{Q^{(a)}} , a=1,\ldots,N \right\}$. In the end the full wave many-body state is obtained by diagonalizing the Hamiltonian in this set of non-orthogonal basis states.

\section{Unitary Correlation Operator Method}

Starting from the realistic Argonne~V18 interaction \cite{wiringa95}
we derive an effective low-momentum interaction using the unitary
correlation operator method (UCOM). The basic idea of the UCOM
approach is to explicitly include short-range central and tensor
correlations into the many-body state by means of a unitary operator
\cite{ucom98,ucom03,ucom10}. The correlation operators can also be
mapped onto the Hamiltonian to define a correlated interaction which not
only contains two-body but also higher order contributions. However,
the UCOM interaction is defined as the two-body part the correlated
Hamiltonian and we neglect higher order terms. The UCOM interaction provides the same phase shifts as the original
interaction but behaves differently in many-body systems. No-core
shell model calculations show that the two-body UCOM interaction gives
a good description of $s$- and light $p$-shell nuclei \cite{ucom10}.

\section{Radiative capture reaction $^3$He($\alpha$,$\gamma$)$^7$Be}

A recent application of the FMD approach is the calculation of the $^3$He($\alpha$,$\gamma$)$^7$Be radiative capture reaction \cite{neff11}. This reaction plays
an important role in the solar proton-proton chains and determines the
production of $^7$Be and $^8$B neutrinos \cite{adelberger98,adelberger11}. This reaction has been studied extensively from the experimental side in recent years \cite{narasingh04,bemmerer06,confortola07,brown07,dileva09}.  However
it is still not possible to reach the low energies relevant for solar burning in experiment.
From the theory side this reaction has been investigated using simple
potential models, where $^3$He and $^4$He are treated as point-like
particles interacting via an effective nucleus-nucleus potential,
e.g., \cite{kim81} or microscopic cluster models, e.g.,
\cite{langanke86,kajino86} where the $^7$Be bound and scattering
states are constructed from microscopic $^3$He and $^4$He clusters
interacting via an effective nucleon-nucleon
interaction. \emph{Ab-initio} calculations using variational Monte
Carlo \cite{nollett01} and no-core shell model wave functions
\cite{navratil07} were used to calculate asymptotic normalization
coefficients for the bound states but relied on potential models for
the scattering phase shifts. 

In the FMD calculation for the $^3$He($\alpha$,$\gamma$)$^7$Be reaction we divide the model
space into two regions. In the external region bound and scattering
states are described by $^3$He and $^4$He clusters in their FMD ground
states. These microscopic wave functions can also be rewritten as
resonating group (RGM) wave functions. The RGM representation allows us to include
boundary conditions
for bound and scattering states by matching to
Whittaker and Coulomb functions at the channel radius ($a$=12~fm). For this we 
employ the microscopic $R$-matrix method developed by the Brussels
group \cite{descouvemont10}. In the interaction region the model space is enlarged by additional FMD
many-body configurations obtained by variation after parity
and angular momentum projection on spin-parity $1/2^+$, $3/2^+$,
$5/2^+$ and $3/2^-$, $1/2^-$, $7/2^-$, $5/2^-$. A constraint on the radius of the intrinsic
states is used to vary the distance between the clusters. 

\subsection{Capture Cross Section}

For energies up to 2.5~MeV only the capture from $S$- and
$D$-wave scattering states into the $3/2^-$ and $1/2^-$ bound states
has to be considered. The addition of polarized configurations is
essential in the FMD framework --- using the frozen configurations only, $^7$Be is
only bound by 200~keV. Including the polarized configurations the $3/2^-$
state is bound by 1.49~MeV and the $1/2^-$ state by 1.31~MeV with
respect to the cluster threshold. The calculated splitting between the two states
is too small compared to the experimental value of 430~keV. This is
related to a deficiency of the two-body interaction -- additional
spin-orbit strength is assumed to be coming from three-body
forces. Fortunately it turns out that the capture cross section
depends strongly on the centroid energy, which is reproduced nicely,
but only very weakly on the splitting between the two bound states. An important test of the ground state wave function is provided by the charge radius.
The calculated value
2.67~fm agrees well with the experimental value of 2.647(17)~fm
\cite{noertershaeuser09}. This is important as the dipole matrix
element and therefore the capture cross section depends on an
accurate description of the tail of the bound state wave functions.
\begin{figure}[t]
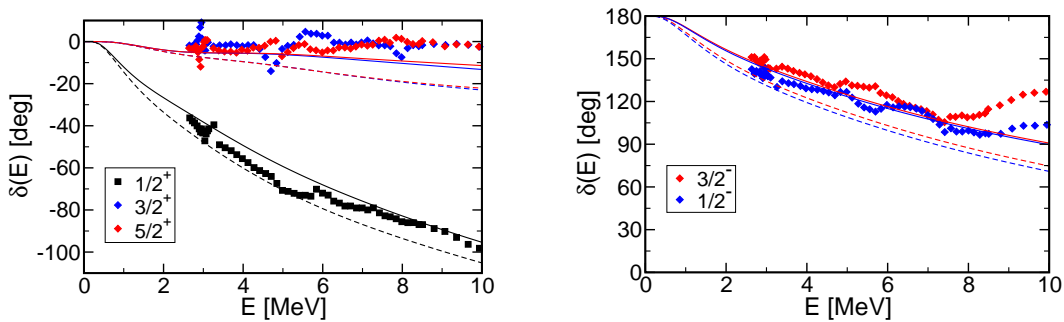

  \centering
  \includegraphics[width=0.40\textwidth]{figs/phaseshifts-sd.eps}\hfil
  \includegraphics[width=0.40\textwidth]{figs/phaseshifts-p.eps}
  \caption{$^4$He-$^3$He scattering phase shifts. Dashed lines show
    results using only frozen configurations, solid lines show results
    with full FMD model space. Left: $S$- and $D$-wave phase
    shifts. Right: $P$-wave phase shifts. Experimental results are
    from \cite{boykin72} and \cite{spiger67}.}
  \label{fig:phaseshifts}
\end{figure}
In Fig.~\ref{fig:phaseshifts} the calculated phase shifts
for the $S$-, $D$- and $P$-waves are compared with the experimental phase shift analyses \cite{boykin72,spiger67}. There is a noticeably change in the results going from the model space containing only the
frozen configurations to the full FMD model space. In case of the $S$-
and $D$-wave phase shifts we find a good agreement with the
experimental data. Also the $P$-wave phase shifts are essentially in
good agreement with experiment, with the exception of the splitting.

\begin{figure}[t]
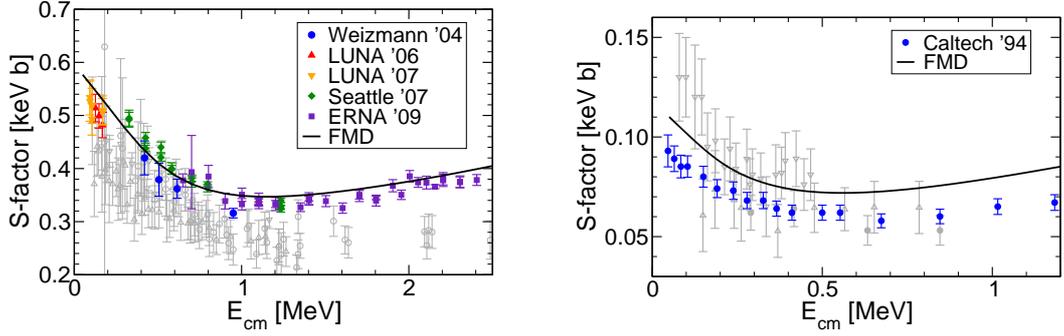

  \centering
  \includegraphics[width=0.40\textwidth]{figs/sfactor-he3-alpha-gamma-be7.eps}\hfil
  \includegraphics[width=0.40\textwidth]{figs/sfactor-h3-alpha-gamma-li7.eps}
  \caption{Left: $S$-factor for the $^3$He($\alpha$,$\gamma$)$^7$Be
    reaction. Recent experimental data
    \cite{narasingh04,bemmerer06,confortola07,brown07,dileva09} are
    shown as colored symbols, older data as gray symbols. Right:
    $S$-factor for the $^3$H($\alpha$,$\gamma$)$^7$Li reaction. Most
    recent data \cite{brune94} is shown as colored symbols, older data
    as gray symbols.}
  \label{fig:sfactor}
\end{figure}

The capture cross section is calculated using the microscopic wave functions of the bound and scattering states.
The result for the total cross section for the $^3$He($\alpha$,$\gamma$)$^7$Be capture is shown in the left part of
Fig.~\ref{fig:sfactor} in form of the astrophysical $S$-factor. It agrees very well with the recent
experimental data both in absolute normalization and in the energy
dependence. Our result for the isospin mirror reaction $^3$H($\alpha$,$\gamma$)$^7$Li is shown on the right hand side of Fig.~\ref{fig:sfactor}. Whereas the energy dependence of the
calculated $S$-factor agrees very well with the data the absolute
cross section is larger then the data by Brune \textit{et al.} by
about 15\%. This is surprising as the FMD results for the $^7$Li bound states and the scattering phase shifts are of similar quality than those for $^7$Be. 

\begin{figure}[b]
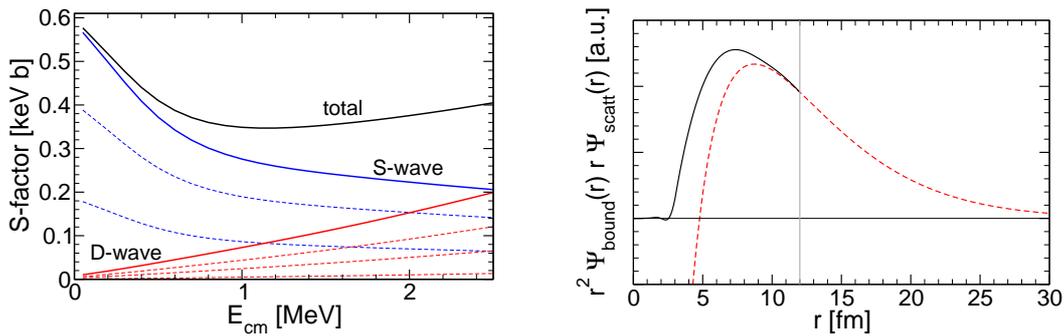

  \centering
  \includegraphics[width=0.40\textwidth]{figs/sfactor-contributions.eps}\hfil
  \includegraphics[width=0.40\textwidth]{figs/dipoleme.eps}
  \caption{Left: $S$- and $D$-wave contributions to the $S$-factor for
    the $^3$He($\alpha$,$\gamma$)$^7$Be reaction. Dashed lines show
    contributions of individual transitions. Right: Dipole strength
    between the $3/2^-$ bound state and the 50~keV $1/2^+$ scattering
    state calculated from the overlap functions (solid line), from the
    Whittaker and Coulomb functions matched at the channel radius
    (dashed line).}
  \label{fig:contributions}
\end{figure}

The total cross section can be decomposed into $S$- and $D$-wave
contributions as shown in the left part of Fig.~\ref{fig:contributions}. If we compare our
results with microscopic cluster model calculations, for example to those by
Kajino \cite{kajino86} we observe that the biggest differences are found
in the $S$-wave contribution. Our results also deviate from the
empirical correlation between radius or quadrupole moment of the
$^7$Be ground state with the $S$-factor at zero energy that was found
in the microscopic cluster model using different phenomenological
interactions.

For a comparison with the cluster model picture we can analyze
the $A$-body bound and scattering state wave functions
in terms of overlap functions. Here the full microscopic wave function
is projected on the subspace of cluster configurations. If one takes
antisymmetrization between the clusters into account properly (by
folding with the square-root of the RGM norm kernel) these overlap
functions can be interpreted as the relative wave functions of
point-like $^3$He and $^4$He clusters. It is interesting that the
overlap functions deviate from the Whittaker and Coulomb functions,
that describe the asymptotic behavior of the cluster motion for bound
and scattering states, up to distances of about 9~fm. These differences
from the asymptotic behavior manifest themselves also in the dipole
strength. The dipole matrix elements calculated from the overlap
functions as shown in the right part of Fig.~\ref{fig:contributions}
agree with the matrix elements from the microscopic wave functions
within 2\%. They have sizable contributions already for distances as
small as 3~fm and deviate significantly from matrix elements
calculated from the asymptotic Whittaker and Coulomb functions to distances of up to
10~fm. This differences indicate that considering this reaction as a simple external capture process is not possible.

\section{Cluster States in $^{12}$C}

\begin{figure}[b]
  \includegraphics[width=0.15\textwidth]{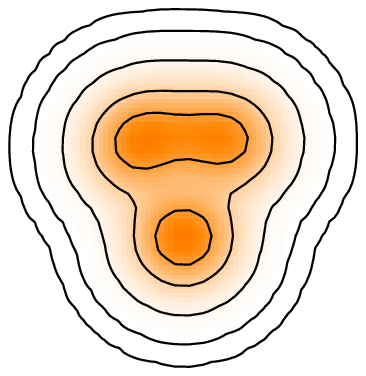}\hfil
  \includegraphics[width=0.15\textwidth]{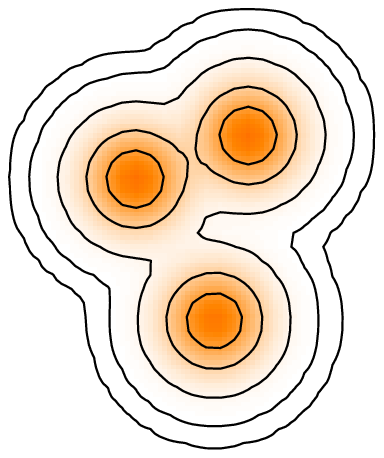}
  \includegraphics[width=0.15\textwidth]{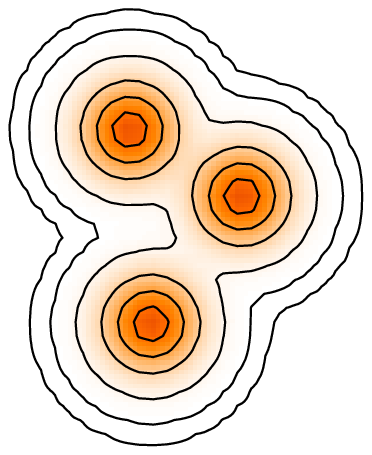}
  \includegraphics[width=0.15\textwidth]{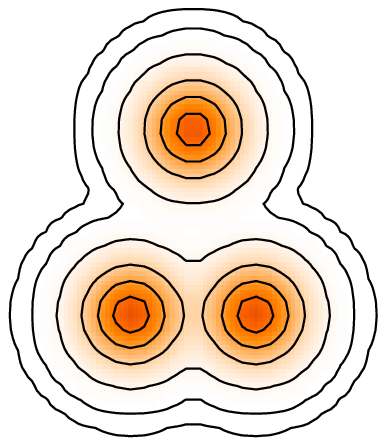}
  \includegraphics[width=0.15\textwidth]{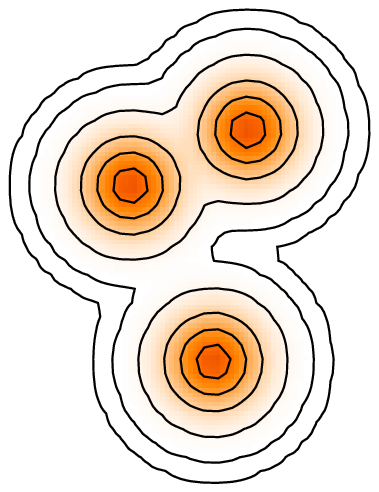}
 \caption{(Left) Intrinsic FMD basis state that has the largest overlap with the ground state. (Right) The four intrinsic FMD basis states that have the largest overlaps with the Hoyle state. The basis states are not orthogonal.}
 \label{fig:intrinsic}
\end{figure}

The structure of the second $0^+$ state in $^{12}$C, the famous Hoyle state, is still one of the hottest topics in nuclear structure. In \cite{Hoyle07} we investigated the structure of the Hoyle state using the FMD approach. The model space consisted of configurations obtained by variation and a full set of three-$\alpha$ configurations. A UCOM interaction with some phenomenological modifications regarding the strength of the spin-orbit force and the saturation properties of the two-body interaction was used in that calculation. We compared the results  with a microscopic cluster model using a phenomenological Volkov interaction. These cluster model calculations reproduced previous results by Kamimura \cite{kamimura81} and are also very close to those obtained by Funaki et al. \cite{funaki03}. We found for both models that the Hoyle state has a very dilute, extended three-$\alpha$ structure. This is illustrated in Fig.~\ref{fig:intrinsic} where we show the intrinsic FMD basis states that have the largest 
overlap with the ground state and the Hoyle state.

We used these wave functions also to calculate the transition form factor from the ground state to the Hoyle state. This transition form factor can be directly compared to electron scattering data \cite{Hoyle07,Hoyle10}. The good agreement of calculation and experiment is a strong confirmation for a spatially extended structure for the Hoyle state.

\subsection{Two-body densities}

Observables like radii and form factors are scalar quantities that provide information about the size of the states but they do not provide direct information about the structure of the states. The old question whether the Hoyle state should be interpreted as a linear chain of $\alpha$-particles, a triangular structure or a gas-like structure can therefore not be answered directly by these experimental observables. There is also the questions of how we should compare the wave functions obtained in different many-body approaches like the cluster model, the no-core shell model or as obtained on the lattice \cite{epelbaum11}.

In case of FMD or the cluster model the individual basis states can be easily interpreted in terms of the intrinsic structure as shown in Fig.~\ref{fig:intrinsic}. However the eigenstates are linear combinations of many basis states and the non-orthogonality of the basis states might question the validity of the obtained picture.

\begin{figure}[t]
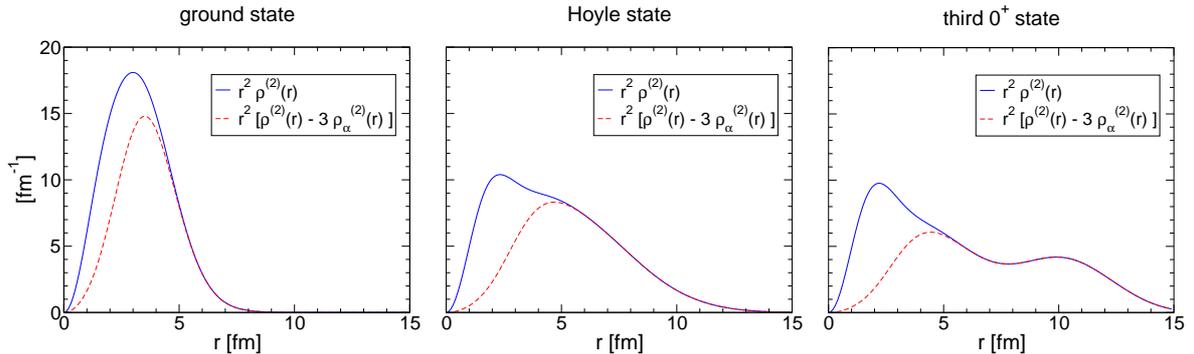

 \centering
 \includegraphics[height=0.29\textwidth]{figs/C12-tbdens-0+1.eps}\hfil
 \includegraphics[height=0.29\textwidth]{figs/C12-tbdens-0+2.eps}\hfil
 \includegraphics[height=0.29\textwidth]{figs/C12-tbdens-0+3.eps}
 \caption{Two-body densities obtained in the cluster model for the $^{12}$C ground state, the Hoyle state, and the third $0^+$ state. The densities are multiplied by $r^2$. Solid lines show the full two-body densities, the dashed lines show the results when the contributions to the two-body densities from inside the $\alpha$-particles is subtracted.}
 \label{fig:twobodydensities}
\end{figure}

To remedy this situation we propose to use two-body densities to analyze the structure of the $^{12}$C eigenstates. In Fig.~\ref{fig:twobodydensities} we show the diagonal part of the two-body density integrated over the center-of-mass coordinates and summed over all spin-isospin channels which can be expressed as
\begin{equation}
 \rho^{(2)}(\vec{r}) = \matrixe{\Psi}{\sum_{i<j} \delta(\op{\vec{r}}_i-\op{\vec{r}}_j - \vec{r})}{\Psi} \: . 
\end{equation}

The two-body density $\rho^{(2)}(\vec{r})$ tells us about the probability to find a pair of nucleons at a given distance $\vec{r}$. In the case of $^{12}$C where we expect an intrinsic $\alpha$-cluster structure the two-body density should directly reflect the correlations between the $\alpha$-clusters. The two-body density is only smeared out by the finite size of the $\alpha$-clusters. In microscopic many-body approaches the spin-isospin dependence of the two-body density could be used to validate the assumption of $\alpha$-clustering. In principle the two-body densities will depend on the interaction used in the many-body approach. However this interaction dependence should only affect the behavior at very short distances. Here we are interested in the long-range correlations. 

In the full two-body density also pairs of nucleons from the same $\alpha$-particle are included. If we assume an $\alpha$-cluster picture the more interesting observable is the probability to find a pair of nucleons from two different $\alpha$-particles. In case of the cluster model one can simply subtract the contributions from within the $\alpha$-particles. These are the dashed curves in Fig.~\ref{fig:twobodydensities}. For the ground state the probability then peaks at about 3.5~fm reflecting the Pauli exclusion principle between the $\alpha$ clusters. For the Hoyle state we still find a single peak in the probability distribution at about 5~fm. This means that the $\alpha$-particles in the Hoyle state are further apart from each other than in the ground state. The probability distribution also extends to much larger distances of more than 10~fm. The fact that there is only a single peak is consistent with the picture of a triangular structure where the distances between all three $\alpha$-particles is 
the same. The density distributions of the third $0^+$ state however shows a two-peak structure, with one peak at about 5~fm and another smaller peak at about 10~fm. This is what one would expect for a chain structure. The $\alpha$-particles at the ends of the chain each see the $\alpha$-particle in the middle of the chain at about 5~fm distance and the one at the other end of the chain at about 10~fm distance.

\subsection{Expansion in oscillator basis}

\begin{figure}[t]
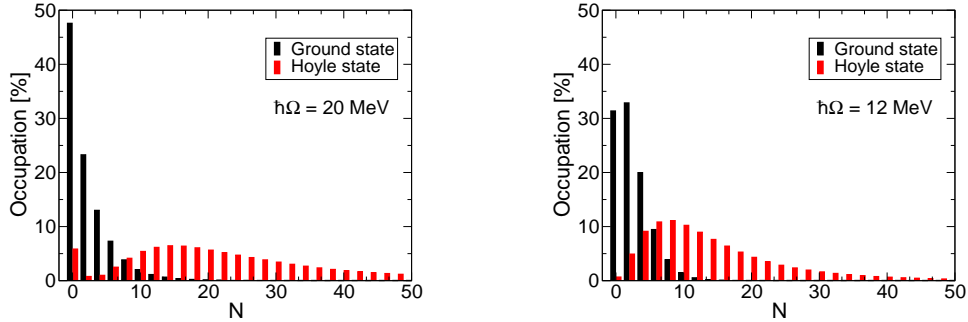

  \centering
  \includegraphics[width=0.34\textwidth]{figs/C12-shelloccupations-20.00-0+.eps}\hfil
  \includegraphics[width=0.34\textwidth]{figs/C12-shelloccupations-12.00-0+.eps}
  \caption{Decomposition of the $^{12}$C ground state and the Hoyle state into $N\hbar\Omega$ components for oscillator constants of 20 MeV (left) and 12 MeV (right).}
  \label{fig:oscillatordecomposition}
\end{figure}

In no-core shell model calculations the second $0^+$ state is found at much too high energies \cite{navratil07}. To illustrate the challenge for the no-core shell model we analyze the cluster model wave functions in the harmonic oscillator basis. Explicitly expanding our wave functions in the harmonic oscillator basis is not feasible. It is however easy to simply count the number of $N\hbar\Omega$ excitations in the wave function. This occupation probability can be calculated by
\begin{equation}
  \mathrm{Occ}(N) = \matrixe{\Psi}{\delta\bigl( 
    \sum_i (\op{H}^{HO}_i/\hbar\Omega - 3/2) - N \bigr)}{\Psi}
\end{equation}
as proposed by Suzuki \textit{et al.} \cite{suzuki96}. The results are shown in Fig.~\ref{fig:oscillatordecomposition} for the cluster model wave functions of the ground state and the Hoyle state. On the left an oscillator parameter $\hbar\Omega$ of 20~MeV was used. That corresponds roughly to the optimal oscillator parameter for the mean-field solution. In case of the ground state the contributions become very small for $N$ larger than 8 or 10. It is therefore not surprising that the NCSM calculations for the ground state can be converged. For the Hoyle state however, the distribution extends over a very large range of $N\hbar\Omega$. It is therefore clear that the Hoyle state can not be converged in NCSM calculations with $N_\mathrm{max} = 8$ or even $10$. The situation looks somewhat better for an oscillator parameter of 12~MeV as shown on the right hand side of Fig.~\ref{fig:oscillatordecomposition}. Here the distribution for the Hoyle state peaks at $N = 8$ and decays much more rapidly with $N$. However,
 standard NCSM calculations will not be able to reach large enough spaces. Maybe approaches like the importance truncated no-core shell model \cite{roth07} or the symmetry adapted no-core shell model \cite{dytrych07} will allow the description of the Hoyle state within the oscillator basis in the future.

\section*{References}
\bibliographystyle{iopart-num}
\bibliography{hites}

\providecommand{\newblock}{}
\begin{thebibliography}{10}
\expandafter\ifx\csname url\endcsname\relax
  \def\url#1{{\tt #1}}\fi
\expandafter\ifx\csname urlprefix\endcsname\relax\def\urlprefix{URL }\fi
\providecommand{\eprint}[2][]{\url{#2}}

\bibitem{wiringa95}
Wiringa R~B, Stoks V~G~J and Schiavilla R 1995 {\em Phys. Rev. C\/} {\bf 51} 38

\bibitem{ucom98}
Feldmeier H, Neff T, Roth R and Schnack J 1998 {\em Nucl. Phys.\/} {\bf A632}
  61

\bibitem{ucom03}
Neff T and Feldmeier H 2003 {\em Nucl. Phys.\/} {\bf A713} 311

\bibitem{ucom10}
Roth R, Neff T and Feldmeier H 2010 {\em Prog. Part. Nucl. Phys.\/} {\bf 65} 50

\bibitem{neff11}
Neff T 2011 {\em Phys. Rev. Lett.\/} {\bf 106} 042502

\bibitem{adelberger98}
Adelberger E~G {\em et~al.\/} 1998 {\em Rev. Mod. Phys.\/} {\bf 70} 1265

\bibitem{adelberger11}
Adelberger E~G {\em et~al.\/} 2011 {\em Rev. Mod. Phys.\/} {\bf 83} 195

\bibitem{narasingh04}
Nara~Singh B~S, Hass M, Nir-El Y and Haquin G 2004 {\em Phys. Rev. Lett.\/}
  {\bf 93} 262503

\bibitem{bemmerer06}
Bemmerer D {\em et~al.\/} 2006 {\em Phys. Rev. Lett.\/} {\bf 97} 122502

\bibitem{confortola07}
Confortola F {\em et~al.\/} 2007 {\em Phys. Rev. C\/} {\bf 75} 065803

\bibitem{brown07}
Brown T~A~D {\em et~al.\/} 2007 {\em Phys. Rev. C\/} {\bf 76} 055801

\bibitem{dileva09}
Di~Leva A {\em et~al.\/} 2009 {\em Phys. Rev. Lett.\/} {\bf 102} 232502

\bibitem{kim81}
Kim B~T, Izumoto T and Nagatani K 1981 {\em Phys. Rev. C\/} {\bf 23} 33

\bibitem{langanke86}
Langanke K 1986 {\em Nucl. Phys.\/} {\bf A457} 351

\bibitem{kajino86}
Kajino T 1986 {\em Nucl. Phys.\/} {\bf A460} 559

\bibitem{nollett01}
Nollett K~M 2001 {\em Phys. Rev.\/} {\bf C63} 054002

\bibitem{navratil07}
Navr{\'a}til P, Gueorguiev V~G, Vary J~P, Ormand W~E and Nogga A 2007 {\em
  Phys. Rev. Lett.\/} {\bf 99} 042501

\bibitem{descouvemont10}
Descouvemont P and Baye D 2010 {\em Rep. Prog. Phys.\/} {\bf 73} 036301

\bibitem{noertershaeuser09}
N{\"o}rtersh{\"a}user W {\em et~al.\/} 2009 {\em Phys. Rev. Lett.\/} {\bf 102}
  062503

\bibitem{boykin72}
Boykin W~R, Baker S~D and Hardy D~M 1972 {\em Nucl. Phys.\/} {\bf A195} 241

\bibitem{spiger67}
Spiger R~J and Tombrello T~A 1967 {\em Phys. Rev.\/} {\bf 163} 964

\bibitem{brune94}
Brune C~R, Kavanagh R~W and Rolfs C 1994 {\em Phys. Rev. C\/} {\bf 50} 2205

\bibitem{Hoyle07}
Chernykh M, Feldmeier H, Neff T, von Neumann-Cosel P and Richter A 2007 {\em
  Phys. Rev. Lett.\/} {\bf 98} 032501

\bibitem{kamimura81}
Kamimura M 1981 {\em Nucl. Phys.\/} {\bf A351} 456

\bibitem{funaki03}
Funaki Y, Tohsaki A, Horiuchi H, Schuck P and R\"opke G 2003 {\em Phys. Rev.
  C\/} {\bf 67} 051306(R)

\bibitem{Hoyle10}
Chernykh M, Feldmeier H, Neff T, von Neumann-Cosel P and Richter A 2010 {\em
  Phys. Rev. Lett.\/} {\bf 105} 022501

\bibitem{epelbaum11}
Epelbaum E, Krebs H, Lee D and Mei{\ss}ner U~G 2011 {\em Phys. Rev. Lett.\/}
  {\bf 106} 192501

\bibitem{suzuki96}
Suzuki Y, Arai K, Ogawa Y and Varga K 1996 {\em Phys. Rev. C\/} {\bf 54} 2073

\bibitem{roth07}
Roth R and Navr{\'a}til P 2007 {\em Phys. Rev. Lett.\/} {\bf 99} 092501

\bibitem{dytrych07}
Dytrych T, Sviratcheva K~D, Bahri C, Draayer J~P and Vary J~P 2007 {\em Phys.
  Rev. Lett.\/} {\bf 98} 162503

\end{thebibliography}

\end{document}